# On the time distribution of Earth's magnetic field reversals

Cosme F. Ponte-Neto[1,*], Andrés R. R. Papa[1,2]

[1]Observatório Nacional, Rua General José Cristino 77, São Cristóvão, Rio de Janeiro, 20921-400 RJ, BRASIL

[2]Instituto de Física, Universidade do Estado do Rio de Janeiro, Rua São Francisco Xavier 524, Maracanã, Rio de Janeiro, 20550-900 RJ, BRASIL

**Abstract**

This paper presents an analysis on the distribution of periods between consecutive reversals of the Earth's magnetic field. The analysis includes the randomness of polarities, whether the data corresponding to different periods belong to a unique distribution and finally, the type of distribution that data obey. It was found that the distribution is a power law (which could be the fingerprint of a critical system as the cause of geomagnetic reversions). For the distribution function a slope value of –1.42 ± 0.19 was found. This value differs about 15% from results obtained when the present considerations are not taken into account and it is considered the main finding.

## Introduction

Geomagnetic reversals (periods during which the geomagnetic field swap hemispheres) are, together with the magnetic storms (because of the immediate effects on man’s activities), the most dramatic events in the magnetic field that we can measure at the Earth’s surface (Merrill, 2004). The time between consecutive geomagnetic reversals has typical values that range from a few tens of thousands of years to around forty millions of years while magnetic storms have durations of approximately two days. They also have different sources, while the magnetic storms are mostly associated to phenomena in the Sun and the terrestrial ionosphere, geomagnetic reversals are associated with changes in the Earth’s dynamo. Towards a deeper understanding of the laws that follow the geomagnetic reversals is devoted this work.

Another common feature of both short period and long period of time phenomena is the appearance of power laws in their relevant distributions (Papa et al., 2006; Seki and Ito, 1993). One of the possible mechanisms that produce power law distributions for, for example, the distribution of times between consecutive periods of great activity, is the mechanism of self-organized criticality or, more specifically, of threshold systems. It is quite remarkable that, phenomena essentially diverse (like magnetic storms and geomagnetic reversals), could be sustained by similar types of mechanisms.

Threshold systems are the base for the behavior of many dissimilar phenomena. They are composed by elements that behave in a special manner: 1) the elements are able to store potential energy up to a given threshold; 2) they are continuously supplied with potential energy; 3) when the accumulate potential energy in an element reach the threshold part of its energy is released to neighbor elements and out of the system; 4) eventually, the energy released to some of the neighbors will be enough to surpass its own threshold; 5) this element will release part of its energy to the neighborhood and out of the system and so on. In this way a single element can spark a long chain reaction that will extinguish only when all the elements are below the threshold. At a first reading the concept of threshold system could appear very abstract, but there are some simple examples that can help in demystifying the concept. Suppose that we locate a block of wood on a surface and attach a spring to it. If we try to move the block by pulling the opposite extreme of the spring initially it will not move. The block will move only when the potential energy accumulated in the spring reach the static friction. In this case the static friction plays the role of threshold. The released energy (as we are considering a single block-spring set there are no neighbors) is composed by the thermal energy (produced by the dynamic friction between the block and the surface) and acoustic energy (the noise that the block produces while sliding on the surface). Actually, models with systems of many spring and interconnected block have served to reproduce some of the main characteristics of earthquakes. The energy has to be supplied at a low rate (compared to the maximum power that the system can dissipate) otherwise there would be no avalanches. In the block-spring example, if we pull the spring very rapidly (i.e. if we introduce energy at a high rate) probably the block will never stop once in movement. It is a usual (non exclusive) signature of self-organized criticality and threshold systems the appearance of power laws

$$f(x) = c \cdot x^{d} \qquad (1)$$

where *x* is the variable, *c* is some proportionality constant, *f(x)* is the distribution of the variable *x* and *d* is the exponent. These concepts will help us in the interpretation of some of the results that we will describe.

Works devoted to the study of the time distribution of geomagnetic reversals include, among many others, an analysis of scaling in the polarity reversal record (Gaffin, 1989), a search for chaos in record (Cortini and Barton, 1994), a critical model for this problem (Seki and Ito, 1993) and more recently, a long-range dependence study in the Cenozoic reversal record (Jonkers, 2003). Gaffin (1989) pointed out that long-term trends and non-stationary characteristics of record could difficult a formal detection of chaos in geomagnetic reversal record. It is our opinion that because of this and also because the low number of reversals, in the work of Gaffin actually, it was pointed out that it would be very difficult to detect in a consistent manner that the geomagnetic reversals present any characteristic at all, without mattering which this characteristic could be (including chaos).

Our study differs from those works in that, we explore the equivalence of both polarizations through some well-known non-parametric test on the reversals time series. We then study the possibility of diverse periods pertain to the same distribution and finally the distribution that geomagnetic reversal effectively follows.

Our work is closer to the one by Jonkers (2003) and in some sense complements it.

**Analysis**

There is some recent evidence (Clement, 2004) on a dependence of the geomagnetic polarity reversals on the site where the analyzed sediments are collected. This can be the fingerprint of higher order (not only dipolar) contributions to the components of the Earth's magnetic field. We have not considered those variations. Another feature that was not considered by us are the detailed variations of the Earth's dipole (Valet et al., 2005). We have just considered polarity inversions. We used the more complete data that we have found (Cande and Kent, 1992, 1995).

In Figure 1 we present the sequences of reversals during the last 120 My. It can be seen a clear difference between the periods 0-40 My and 40-80 My, before the great Cretaceous isochrone. Our intuitive reasoning can be further supported by some evidences of tectonic changes experienced by the Earth at the same epoch (around 40 My ago) that could have influenced the dynamo system: the change in direction of growth of the Hawaiian archipelago. Those are the reasons to study separately, at least initially, both periods.

We wonder now, are both polarizations in each of the periods equally probably? If both polarizations are equivalent this is a useful fact from the statistical point of view. Instead of two small samples we have a single and larger one. At the same time, the equivalence might be pointing to an almost inexistence of tectonic influence on the reversal rate because the Earth has a defined rotation direction (although the rotation is considered a necessary condition). On the other hand, an almost inexistent influence (or very small influence) is compatible with the requirement of self-organized criticality and threshold systems of a small energy deliver rate. However, see below.

We have implemented a non-parametric sequence *u* test. To do so we have taken the shortest interval in each period between consecutive geomagnetic reversals as a trial (0.01 My and 0.044 My for 0-40 and 40-80 My periods, respectively). We normalized to this value the rest of the reversals in each period. The result (rounded) was taken as a sequence of identical consecutive trials for that polarization. In this way we obtained a sequence of the type (*N* means normal and *R* means reverse polarization) “*NNNRRNNNNNRRRNRNNN* …”, over which we implemented the test. For the period 0-40 My, that includes around 140 reversals, it was obtained that both polarizations are almost identically probable (1966 trials in one polarization against 1985 in the opposite one). On the other hand, for the period 40-80 My, that contains only 40 reversal, the result was no so good: for one polarization we obtained 632 trials while for the opposite one only 353 trials. There are two possible explanations for this fact: there was some factor that favored a polarization over the other (of tectonic nature, for example) or the sample is not large enough to avoid fluctuations (note that the number of reversals in the 40-80My period is around 25% the number in the 0-40 My period). We will assume that the second explanation is the actual one. There are no reasons to believe that the mechanism producing the reversals has changed in nature. Consequently, for each of the periods both reversal polarities have been considered as a single sample. The other relevant result that we can extract from the trials is that we must reject the null hypotheses $H_0$ of randomness almost with a 100% confidence. This result coincides with a previous one (Jonkers, 2003), but to arrive to that conclusion there were used specific methods (aggregate variance and absolute value) devised for long-range-dependences studies.

A natural question that arises is, do both periods correspond to the same distribution? Before trying to answer this question let us make some considerations on distribution functions. From a “classical” point of view, belong-to-the-same-distribution means to have similar means and standard deviations (this assertion includes many distribution function types like gaussians, lorentzians, etc.). When we work with power-law distribution functions special cares have to be taken because the distribution are endless. This can be easily seen in a log-log plot. In this type of plot the distribution takes the form of a straight line. So, belong-to-the-same-distribution could well mean that both data sets fit the same straight line but in different intervals. To try answering the question we separately present in Figure 2 the frequency distribution of reversals for the two periods using log-log scale and logarithmic bins. Both distributions present approximately a top-of-a-bell shape but with maximum at different values of time. Logarithmic bins constructions have the property of converting exact (functional) power-law distribution functions with exponent $d$, in power laws with exponent $d$+1. At the same time, if there is a reasonable number of data, they produce best quality (soft) curves because they average (integrate) over increasing windows. From Figure 2 it can be seen that for small time periods both curves initially grow (which means that the distribution, if following some power law, presents an exponent $d \geq 0$). For the highest values (again, if following a power law) the exponent is $d < -1$ (because for $d = -1$ the logarithmic bin plots would be constant values). However, the number of points is not large enough for more accurate predictions on the exponents from this type of graph. From the shape we deduce that they follow the same law (following previous works we believe to be a power law with a unique slope). Supposing that they effectively follow a power law then they also should rest approximately on a single straight line: fortunately, we should not be worried with the weight of each of the periods because the time (which means, statistical weight) is approximately the same for both periods. However, this poses a problem to construct a single histogram with both periods (i.e., to consider both periods as part of a single sample): the middle values could be counted twice while the extremes just once. In order to compare considering both periods as a single sample or as two separate samples we constructed the frequency distribution from the whole period from 0 to 80 My. Figure 3 shows the result. A linear fit to the data gave a value of $-1.64 \pm 0.24$ for the slope. We have then constructed independently the frequency distribution for each of the periods and represented them in a single plot. The result is shown in Figure 4. The slope of the linear fit to both data takes a value $-1.42 \pm 0.19$, well apart from the result that we have found when not taking into account our present considerations (however, within the error interval). The most accepted value for this slope is $\sim -1.5$, near the average of the two that we have found.

Self-organized systems have no a typical time scale nor a typical length scale (and the behavior in time and in space are closely related, both are fractals). The unique relevant length is the system size. The same model system with different sizes gives results that depend on the size in the way we explain now. As an example let us take a simple model for the brain (Papa and da Silva, 1997). If we simulate the model using 1024 elements we will obtain power law distributions for the first return time with a slope of –1.58. If we now use 4096 elements we will obtain the same power law dependence with the same slope. The difference between both cases is that while for the case of 1024 elements we obtain “clean” power laws for about two decades, when we use 4096 we can extend this interval to around four decades. So, different intervals in the same power law could indicate different sizes of activity regions for geomagnetic currents. Besides the fact of having small samples, this is a factor that could partially explain why the distributions go in the form of a power law to lower or higher values, to the right. Another factor that can limit the extension of power laws by the left (small values) is the rate at which energy is delivered to the system. It is a threshold (can not be confused with the threshold mentioned at the introductory section) for the smaller avalanches that exist and can be observed. In this way it should increase the average value between consecutive avalanches or, in other words, will cause an increase in first return times (the equivalent of reversals for the present problem).

## Conclusions

Using classical statistical analysis we have excluded the possibility of reversals be a random process (or the result of a random process), conclusion that coincides with previous ones demonstrated through different methods. From the period 0-40 My (and in a less degree, from the period 40-80 My), where the probability of both polarities was almost identical, we can conclude that the influence of the geodynamo on reversals is null or very small. This fact is compatible with the necessity for self-organized criticality and threshold systems of a small energy release rate. From our results we can also conclude that the existence of power laws in the time distribution of geomagnetic reversals is a probable fact. The existence of power laws can be the result of many mechanisms. So, our results do not demonstrate the existence of a critically self-organized (or even a simple critical) system as the source for geomagnetic reversals but they are compatible with these possibilities. The value of –1.42 for the slope of the distribution function is an original finding and needs further confirmation by other authors. Modeling of the source system for reversals is an exciting problem. Some works are currently running with this aim and will be published elsewhere.

## Acknowledgements

The authors sincerely acknowledge partial financial support from FAPERJ (Rio de Janeiro Founding Agency) and CNPq (Brazilian Founding Agency).

**Figure Captions**

Figure 1.- Representation of geomagnetic reversals from 120 My ago to our days. We arbitrarily have assumed –1 as the current polarization.

Figure 2.- Log-log plot of the distributions of intervals between consecutive reversals for the periods from 0 to 40 My (squares) and from 40 to 80 My (circles). We have used logarithmic bins of size $0.015 \times 2^n$ My, where n=0, 1, 2, 3, 4, 5, 6 and 7. To highlight the similarity between both curves they were normalized to have approximately the same height.

Figure 3.- Frequency distribution for the period from 0 to 80 My. The bold straight line is a linear fit to the data. It has a slope $-1.64 \pm 0.24$.

Figure 4.- Frequency distributions for the periods from 0 to 40 and from 40 to 80 My. From left to right the points of numbers 1, 3, 4, and 6 belong to the period from 0 to 40 My. Points with number 2, 5, 7, 8 and 9 belong to the 40 – 80 My period. The bold straight line is a simultaneous linear fit to both data. It has a slope value of $-1.42 \pm 0.19$.

Figure 1

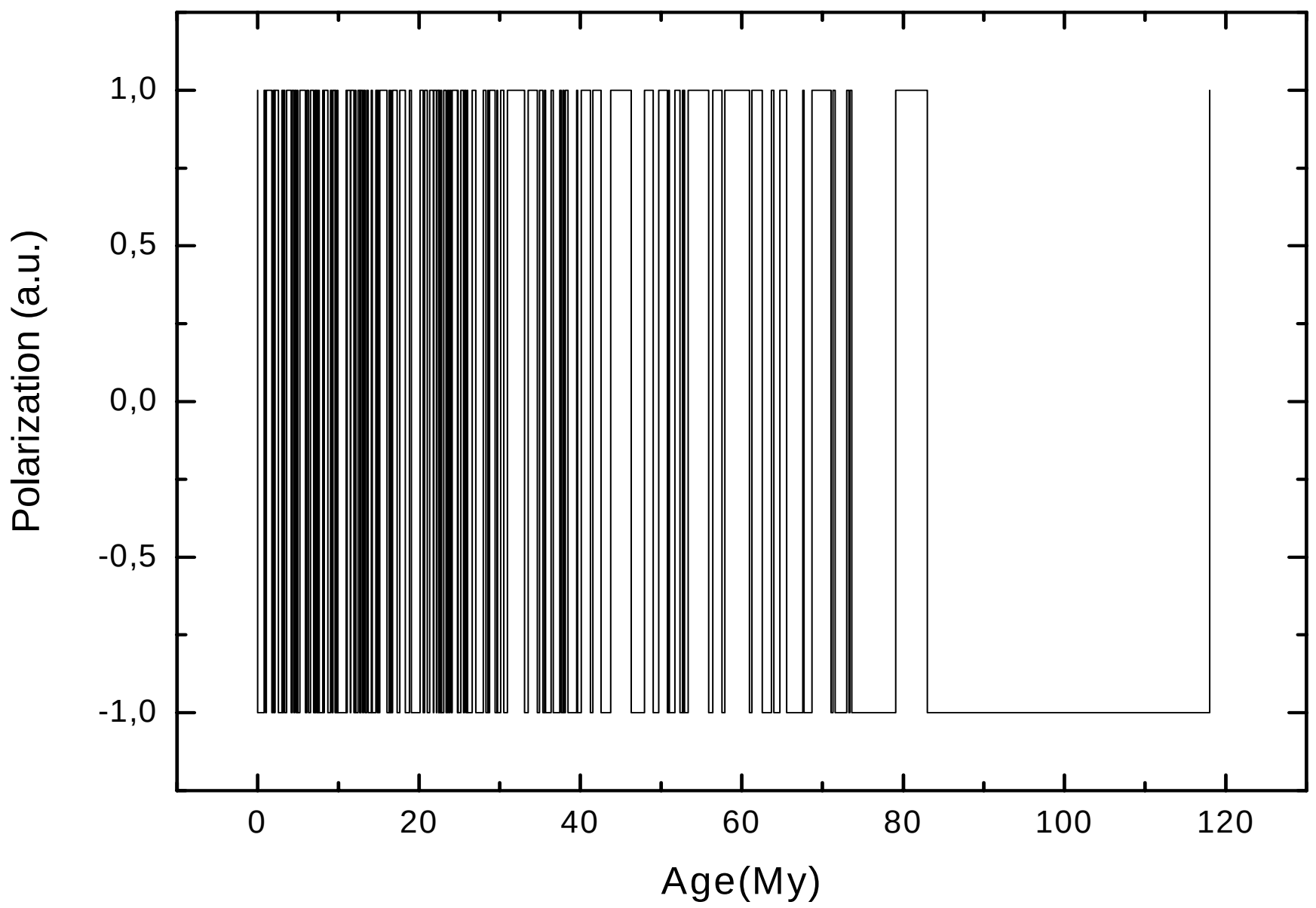

Figure 2

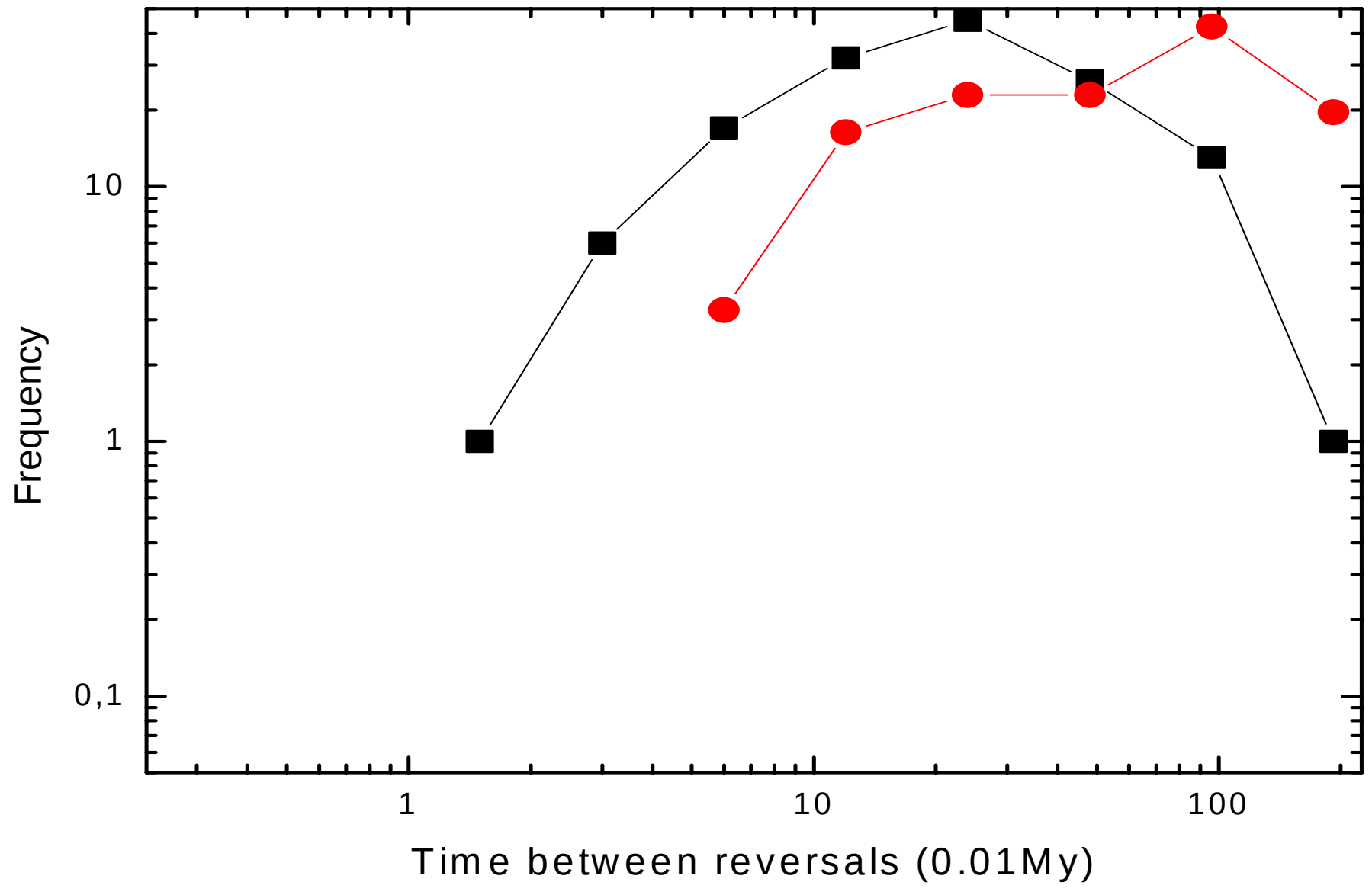

Figure 3

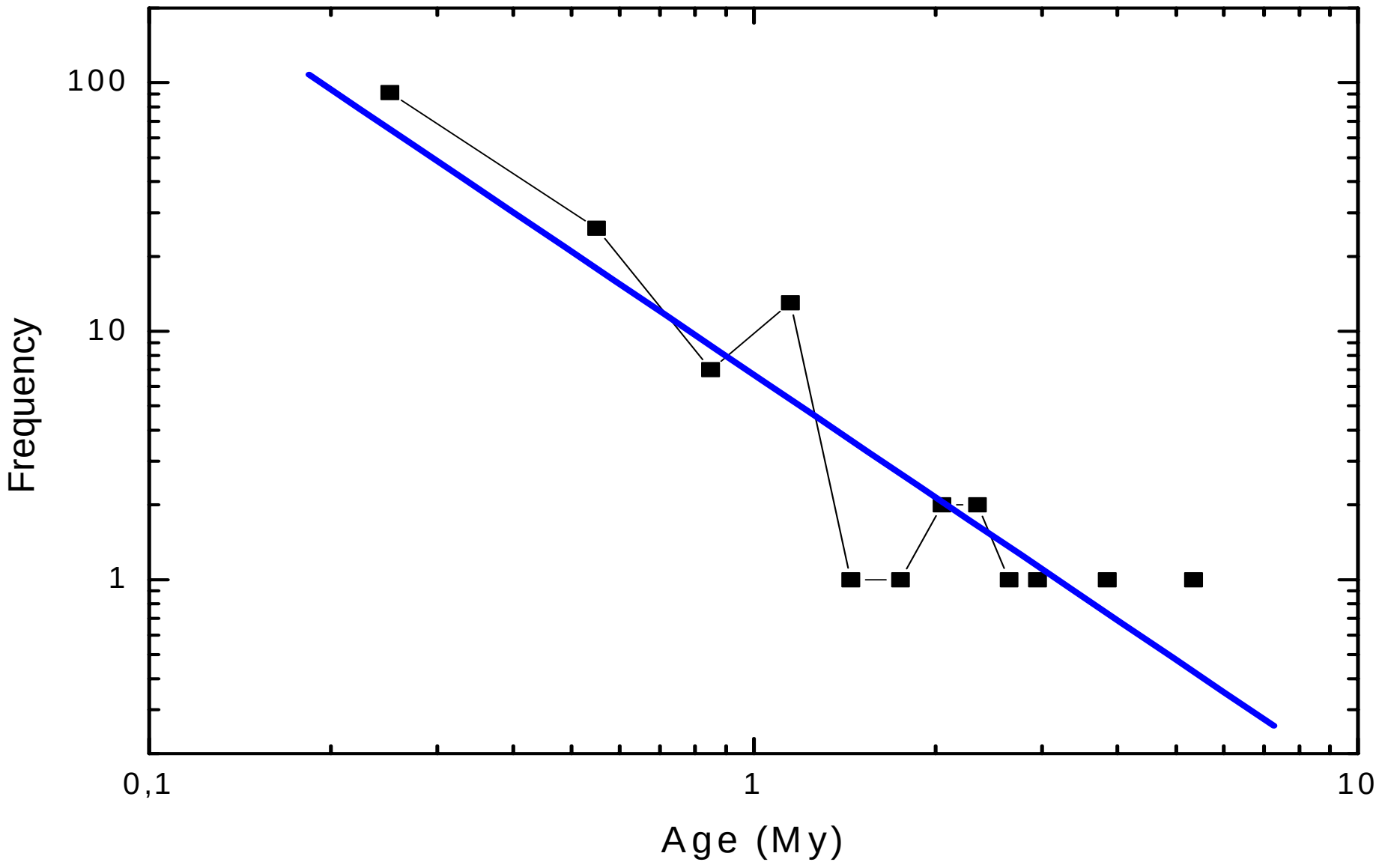

Figure 4

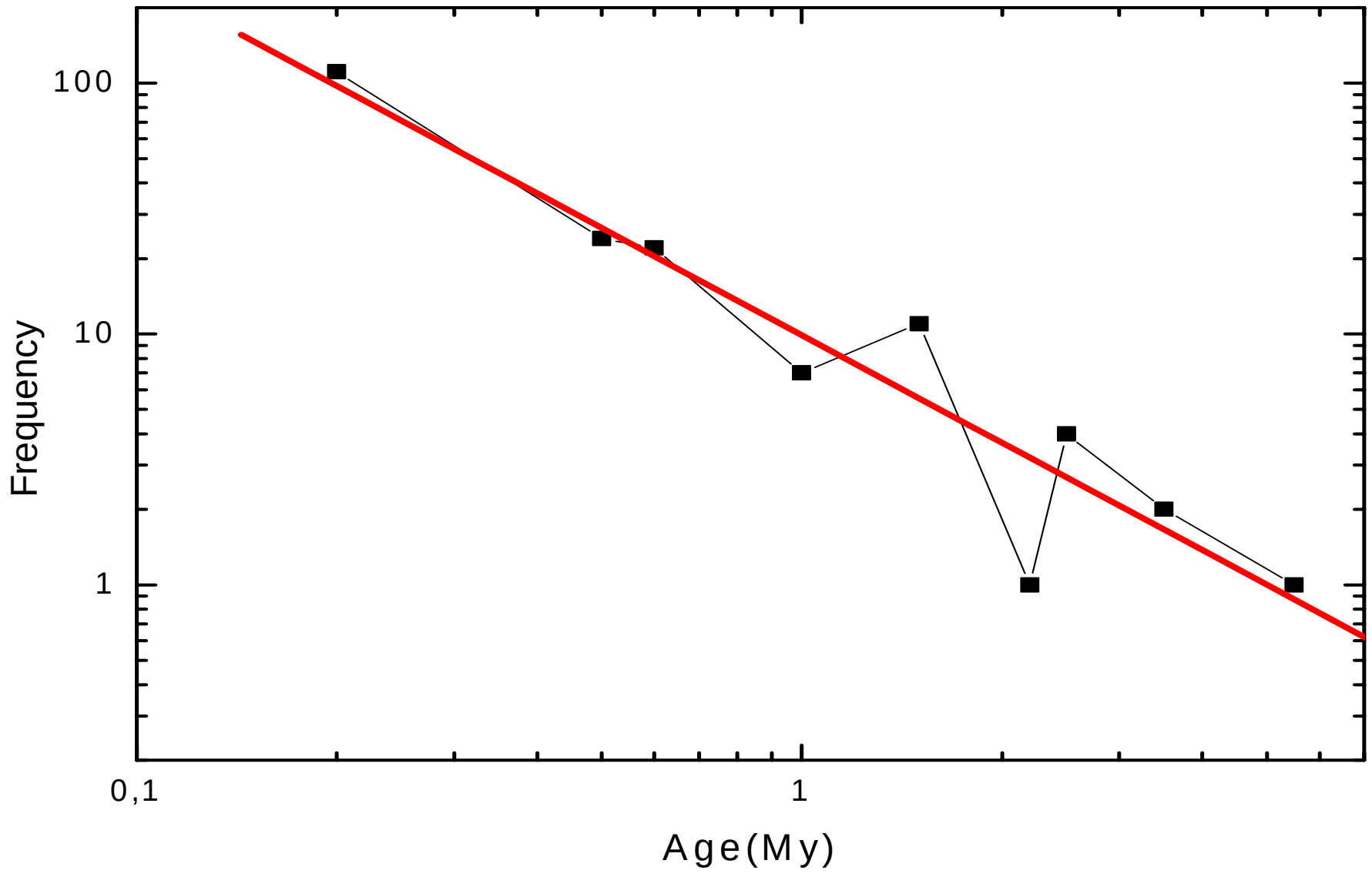